\documentclass{esannV2}
\usepackage{graphicx}
\usepackage[latin1]{inputenc}
\usepackage{amssymb,amsmath,array}
\usepackage{fancyhdr}
\usepackage{multirow}
\usepackage[normalem]{ulem}
\useunder{\uline}{\ul}{}
\usepackage{float}

\usepackage{siunitx}
\usepackage[a4paper,top=3.5cm,bottom=3.5cm,left=3.6cm,right=3.6cm,headsep=20pt, headheight=1cm]{geometry}

\fancypagestyle{title}{%
  \setlength{\headheight}{62pt}%
  \fancyhf{}
  \fancyhead[C]{\fontsize{7}{5} \selectfont ESANN 2023 proceedings, European Symposium on Artificial Neural Networks, Computational Intelligence and
Machine Learning. Bruges (Belgium) and online event, 4-6 October 2023, i6doc.com publ., ISBN 978-2-87587-088-9.
Available from http://www.i6doc.com/en/.}

}

\begin{document}
\title{Multimodal Approach for Harmonized System Code Prediction}

\author{Otmane Amel$^1$\thanks{These authors contributed equally to this work} , S\'edrick Stassin$^1$$^*$, Sidi Ahmed Mahmoudi$^1$ and Xavier Siebert$^1$
\thanks{The authors thank the support of Infortech institute and the E-origin project funded by the Walloon Region within the pole of logistics in Wallonia.}
%
\vspace{.3cm}\\
%
University of Mons - ILIA Unit \\
Mons - Belgium
%
}

\maketitle
\thispagestyle{title}

\begin{abstract} 
The rapid growth of e-commerce has placed considerable pressure on customs representatives, prompting advanced methods. In tackling this, Artificial intelligence (AI) systems have emerged as a promising approach to minimize the risks faced. Given that the Harmonized System (HS) code is a crucial element for an accurate customs declaration, we propose a novel multimodal HS code prediction approach using deep learning models exploiting both image and text features obtained through the customs declaration combined with e-commerce platform information. We evaluated two early fusion methods and introduced our MultConcat fusion method. To the best of our knowledge, few studies analyze the feature-level combination of text and image in the state-of-the-art for HS code prediction, which heightens interest in our paper and its findings. The experimental results prove the effectiveness of our approach and fusion method with a top-3 and top-5 accuracy of 93.5\% and 98.2\% respectively.

\end{abstract}

\section{Introduction}
Repeated legislative changes, along with the massive increase in e-commerce flows, have significantly altered declaration procedures and raised the risks that customs representatives must bear. Today, there is a clear need to reduce the declarative risks in the customs field.
The tariff classification of products is a crucial component in customs declarations. Customs representatives rely on Harmonized System (HS) codes to classify the products for declaration based on information provided by their clients.
The HS codes provide a hierarchical categorization of the products using 6 digits, with each product assumed to belong to a specific category. It starts with two digits (called section or HS2) that define broad categories of products and continues up to six international digits (sub-heading or HS6) which describes a specific product with precision\footnote{Example of the Belgian governmental free access
database of HS code nomenclature: https://eservices.minfin.fgov.be/extTariffBrowser/browseNomen.xhtml?suffix=80\&lang=EN}. Additional digits are added to classify products down to the national level, which determines the exact tax rate for a product. Mistakes made during the customs declaration process can lead to incorrect HS codes being entered into declarations. As a result, an incorrect tax rate may be paid at the end. This emphasizes the importance of the HS code in customs declarations.
Therefore, artificial intelligence (AI) systems that seek to verify the classification of goods are perfectly suited to assist users such as customs representatives.
In this paper, we propose a multimodal model using different textual features as well as images, retrieved from customs declarations coupled with information extracted from the e-commerce website related to the purchase of the product. This model classifies six-digit HS codes (HS6) with an accuracy of top-3 and top-5 accuracy of 93.5\% and 98.2\% respectively. The remainder of this paper is structured as follows. Section \ref{sec:relwork} presents the related works. Section \ref{sec:method} introduces the proposed approach. Section \ref{sec:experiments} presents the experimental results. Finally, Section \ref{sec:conclusion} presents a conclusion and gives some perspectives for future work.
\section{Related Works} \label{sec:relwork}
In this Section, we present works that combine text and image modalities in the e-commerce field. Then, we focus in particular on the papers related to the HS code prediction problem itself. However, the necessity for labeled data and the fact that datasets used in research publications come from private sources and remain secret are recurring challenges for results comparison. As a result, it is hard to compare results in the field accurately.
The work of Zahavy et al.~\cite{zahavy2018picture} uses a dataset of 1.2 million items collected from Walmart.com containing the image, the title as well as the shelf (product categories) to predict between 2890 possibilities. Their experiments proved the effectiveness of late fusion using a learned policy based on class probabilities to combine the convolutional neural networks (CNN) decisions reaching 70.2\% for text, 56.1\% for image and 71.8\% for their combination. 
Chen et al.~\cite{chen2021multimodal} employed a dataset of 500,000 e-commerce products to predict the category based on Japanese titles and images. Their best result ranged between 72.9 and 81.5\% according to the categories predicted, obtained using vision transformers~\cite{dosovitskiy2020image} (ViT) and Japanese BERT~\cite{devlin2018bert} with the use of cross-modal attention as an early fusion method for the modalities.
In the field of multimodal HS code prediction, Turhan et al.~\cite{turhan2015visual} proposed a topic modeling approach based on the product description and image. They offer the user the most similar images with the corresponding HS code to facilitate their choice. Their approach achieves a top-10 accuracy of 87.1\% and 78.9\% for HS4 and HS6, respectively.
Another work by Li and Li~\cite{li2019customs} uses a separately trained  text and image CNN. By grouping six similar HS codes into four classes with 2500 data each, they obtain an accuracy of 93.4\% for the text model, 76.9\% for the image model, and 93.9\% for the combination of the two using a late fusion based on weights calculated using the model accuracies. To the best of our knowledge, few studies analyze the combination of text and image in the state-of-the-art, which heightens  interest in our paper and its findings. 
We differ from the previous works in the following way. 1) We study the combination of image and multiple text modalities to enhance HS code prediction. 2) We conduct a comparative analysis of fusion methods at the feature level (early fusion) and we propose our improved early fusion method MultConcat using arithmetic operations inspired by~\cite{rodrigues4292754multimodal}. 3) We examine the impact of the visual modality through a comparative analysis of two transformer-based and CNN feature extractors.


\section{Methodology} \label{sec:method}

\begin{figure}[!htbp]
\centering
\includegraphics[width=0.695\textwidth]
{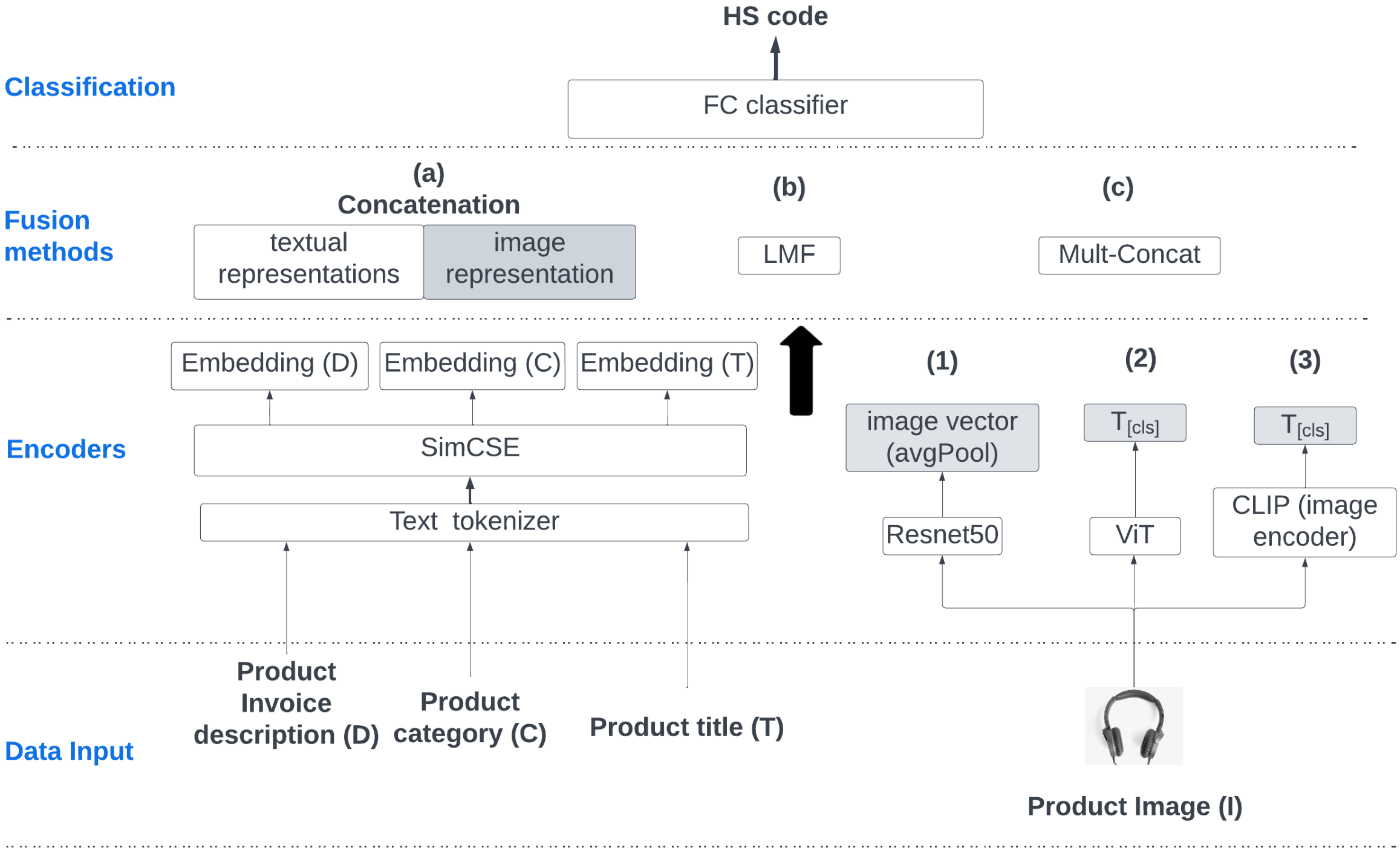}
\caption{Proposed multimodal architecture for HS Code prediction.
}
\label{fig:modelarchitecture}
\end{figure}

We present our proposed architecture for HS code prediction as depicted in Figure~\ref{fig:modelarchitecture}. In this work, two types of modalities are available: text and image. The visual modality consists of product image denoted $I$, and the following are textual modalities: invoice description denoted $D$ from the customs declaration, product title denoted $T$, and product category denoted $C$. The $I$, $T$, and $C$ features are extracted from the e-commerce platform. We employed these encoders for their renowned feature extraction capabilities \cite{zahavy2018picture,chen2021multimodal}: Resnet50 \cite{he2016deep}, ViT~\cite{dosovitskiy2020image}, and CLIP's image encoder~\cite{radford2021learning}. To obtain the final representation of the features, we extracted the 2048 intermediate features from the \textit{avgpool} layer of Resnet50 and used the classification token $\text{T}_{cls}$ for the two transformers models. The textual modalities are fed to the pre-trained model SimCSE~\cite{gao2021simcse}, a widely used sentence embedding extractor of 768 dimensions. Next, we merge the modalities with different early fusion methods such as a simple concatenation \cite{ramachandram2017deep} (Concat) of each modality representation ${M}_i$ or a multimodal low-rank tensor fusion (LMF) \cite{liu2018efficient}. Inspired from \cite{rodrigues4292754multimodal} we propose an enhanced multiplication fusion approach called MultConcat. After projecting linearly each of the $N$ modalities $M_i$ in the same vector space through a hidden layer  resulting in $out_i$ (see Eq.~\ref{eq1} where $W_i$ and $b_i$ are learnable parameters), MultConcat is obtained based on the concatenation $\Vert$ of two terms (see Eq.~\ref{eq2}): a concatenated representation of each $out_i$ called $C$, and an element-wise multiplication $\odot$ (or called Hadamard product) of each $out_i$ called $Z$ (see Eq.~\ref{eq3}). 

\begin{equation}
    out_i = \text{ReLU}(W_i \times M_i + b_i)
    \label{eq1}
\end{equation}

\begin{equation}
    \text{MultConcat} = C \mathbin\Vert Z
    \label{eq2}
\end{equation}

\begin{equation}
 \begin{aligned}
 C = \|_{i=1}^N out_i  \qquad \qquad
 Z = \odot _{i=1}^{N} out_i
 \end{aligned}
 \label{eq3}
\end{equation}

Finally, the resulting multimodal representation vector is fed to a one-layer classifier for HS code prediction. 

\section{Experiments} \label{sec:experiments}

\subsection{Dataset}
The dataset consists of 2144 customs declarations provided by our project partner e-Origin\footnote{https://eorigin.eu/}, having a total of 16 distinct HS6 codes along with customs declaration information as well as additional ones provided by the marketplace from where the goods originated. The database underwent a preprocessing step, during which we eliminated punctuation, special characters, and digit numbers from the textual columns. Subsequently, we observed miswritten or concatenated words that needed unique attention. To address this issue, we utilized a blend of word-separating tools from the WordSegment library\footnote{https://grantjenks.com/docs/wordsegment/} and a Python-based spell-checker\footnote{https://pyspellchecker.readthedocs.io/}. In addition, product images were normalized and resized to fit the image encoders' requirements. Finally, the dataset was split into train, test, and validation sets with a ratio of 80\%, 10\%, and 10\% respectively. 

\subsection{Setup}
The comparative analysis was performed using SimCSE  public checkpoint\footnote{\textit{sup-simcse-bert-base-uncased} weights from https://github.com/princeton-nlp/SimCSE}, in addition to \textit{vit-base-patch16-224-in21k} weights for ViT\footnote{https://huggingface.co/google/vit-base-patch16-224-in21k}, and \textit{ViT-B/32} for CLIP image encoder\footnote{https://huggingface.co/sentence-transformers/clip-ViT-B-32}. For fusion methods, we took the implementation of LMF using the framework Multibench \cite{liang2021multibench} with a decomposition rank factor of 16. Our models were trained and evaluated on GPU resources using the PyTorch framework. The Adam optimizer \cite{kingma2014adam} was used with a $1e^{-4}$ learning rate, a total of 100 epochs, and an early stopping set to 10 epochs.
\subsection{Discussion}
In Table \ref{tab:result} below we provide our comparative results by varying the fusion methods and the image encoders. The top-$k$ accuracies are achieved on the test sets. The goal of this analysis is to assess whether multimodal architectures based on early fusion methods help to boost the HS recommendation results compared to baseline models represented in the bottom part of the table. One can notice an improvement by a large margin with the multimodal networks except those employing the LMF fusion technique, this might be interpreted by a loss of cross-modal information during the decomposition phase. Additionally, our proposed fusion method MultConcat comes ahead of the other fusion methods since it gives overall better performance than simple concatenation or LMF. Moreover, different image encoders were employed for the visual modality I, and we notice that ViT gives the highest top-1 accuracy  of 65.3\%. However, ResNet50 yields better results in terms of top-3 and top-5 accuracy metrics. Although by a low margin compared to ViT, this suggests that ResNet50 encoders enhance the flexibility of our multimodal architectures when employed for recommendation tasks. This is the desired characteristic of the model deployed during inference, as customs representatives need the flexibility to select the correct HS code between multiple choices. It is worth noting that adding the visual modality $I$ to the initial invoice description $D$ improves the top-1 accuracy by 8.2\% compared to the unimodal approach. When set against the combined textual modalities ($T$, $D$, $C$), this improvement is marginal (0.6\%). The limited enhancement might result from noisy image products that aren't directly related to the textual information, hindering overall performance.

\begin{table}[!htbp]
\centering
\begin{tabular}{lllllll}
\hline
\multicolumn{1}{|l|}{\multirow{2}{*}{\textbf{\begin{tabular}[c]{@{}l@{}}Fusion \\ method\end{tabular}}}} & \multicolumn{2}{c|}{\textbf{Encoder}}                                                         & \multicolumn{1}{l|}{\multirow{2}{*}{\textbf{Modality}}} & \multicolumn{3}{c|}{\textbf{Top-k}}                                                                             \\ \cline{2-3} \cline{5-7} 
\multicolumn{1}{|l|}{}                                                                                   & \multicolumn{1}{c|}{\textbf{Image}}            & \multicolumn{1}{c|}{\textbf{Text}}           & \multicolumn{1}{l|}{}                                   & \multicolumn{1}{c|}{\textbf{k=1}}   & \multicolumn{1}{c|}{\textbf{k=3}}   & \multicolumn{1}{c|}{\textbf{k=5}}   \\ \hline
\multicolumn{1}{|l|}{MultConcat}                                                                         & \multicolumn{1}{l|}{\multirow{3}{*}{ViT}}      & \multicolumn{1}{l|}{\multirow{3}{*}{SimCSE}} & \multicolumn{1}{l|}{I,T,D,C}                            & \multicolumn{1}{l|}{\textbf{0.653}} & \multicolumn{1}{l|}{{0.929}}    & \multicolumn{1}{l|}{{\underline{0.977}}}    \\ \cline{1-1} \cline{4-7} 
\multicolumn{1}{|l|}{Concat}                                                                             & \multicolumn{1}{l|}{}                          & \multicolumn{1}{l|}{}                        & \multicolumn{1}{l|}{I,T,D,C}                            & \multicolumn{1}{l|}{0.624}          & \multicolumn{1}{l|}{{0.924}}    & \multicolumn{1}{l|}{{\underline{0.977}}}    \\ \cline{1-1} \cline{4-7} 
\multicolumn{1}{|l|}{LMF}                                                                                & \multicolumn{1}{l|}{}                          & \multicolumn{1}{l|}{}                        & \multicolumn{1}{l|}{I,T,D,C}                            & \multicolumn{1}{l|}{0.088}          & \multicolumn{1}{l|}{0.188}          & \multicolumn{1}{l|}{0.347}          \\ \hline
\multicolumn{1}{|l|}{MultConcat}                                                                         & \multicolumn{1}{l|}{\multirow{3}{*}{ResNet50}} & \multicolumn{1}{l|}{\multirow{3}{*}{SimCSE}} & \multicolumn{1}{l|}{I,T,D,C}                            & \multicolumn{1}{l|}{0.612}          & \multicolumn{1}{l|}{\textbf{0.935}} & \multicolumn{1}{l|}{\textbf{0.982}} \\ \cline{1-1} \cline{4-7} 
\multicolumn{1}{|l|}{Concat}                                                                             & \multicolumn{1}{l|}{}                          & \multicolumn{1}{l|}{}                        & \multicolumn{1}{l|}{I,T,D,C}                            & \multicolumn{1}{l|}{0.571}          & \multicolumn{1}{l|}{{ 0.924}}    & \multicolumn{1}{l|}{\underline{0.977}}          \\ \cline{1-1} \cline{4-7} 
\multicolumn{1}{|l|}{LMF}                                                                                & \multicolumn{1}{l|}{}                          & \multicolumn{1}{l|}{}                        & \multicolumn{1}{l|}{I,T,D,C}                            & \multicolumn{1}{l|}{0.047}          & \multicolumn{1}{l|}{0.182}          & \multicolumn{1}{l|}{0.241}          \\ \hline
\multicolumn{1}{|l|}{MultConcat}                                                                         & \multicolumn{1}{l|}{\multirow{3}{*}{CLIP}}     & \multicolumn{1}{l|}{\multirow{3}{*}{SimCSE}} & \multicolumn{1}{l|}{I,T,D,C}                            & \multicolumn{1}{l|}{{0.629}}    & \multicolumn{1}{l|}{0.918}          & \multicolumn{1}{l|}{\underline{0.977}}          \\ \cline{1-1} \cline{4-7} 
\multicolumn{1}{|l|}{Concat}                                                                             & \multicolumn{1}{l|}{}                          & \multicolumn{1}{l|}{}                        & \multicolumn{1}{l|}{I,T,D,C}                            & \multicolumn{1}{l|}{0.624}          & \multicolumn{1}{l|}{0.924}          & \multicolumn{1}{l|}{\underline{0.977}}          \\ \cline{1-1} \cline{4-7} 
\multicolumn{1}{|l|}{LMF}                                                                                & \multicolumn{1}{l|}{}                          & \multicolumn{1}{l|}{}                        & \multicolumn{1}{l|}{I,T,D,C}                            & \multicolumn{1}{l|}{0.277}          & \multicolumn{1}{l|}{0.359}          & \multicolumn{1}{l|}{0.477}          \\ \hline
\multicolumn{1}{|l|}{MultConcat}                                                                                  & \multicolumn{1}{l|}{/}                         & \multicolumn{1}{l|}{SimCSE}                  & \multicolumn{1}{l|}{T,D,C}                                  & \multicolumn{1}{l|}{\underline{0.647} }           & \multicolumn{1}{l|}{\underline{0.930}}          & \multicolumn{1}{l|}{0.970}          \\ \hline
\multicolumn{1}{|l|}{MultConcat}                                                                                  & \multicolumn{1}{l|}{RestNet50}                         & \multicolumn{1}{l|}{SimCSE}                  & \multicolumn{1}{l|}{I,D}                                  & \multicolumn{1}{l|}{0.582}           & \multicolumn{1}{l|}{0.870}          & \multicolumn{1}{l|}{0.924}          \\ \hline
\multicolumn{7}{c}{baseline (unimodal models)}                                                                                                                                                                                                                                                                                                                                       \\ \hline
\multicolumn{1}{|l|}{/}                                                                                  & \multicolumn{1}{l|}{/}                         & \multicolumn{1}{l|}{SimCSE}                  & \multicolumn{1}{l|}{D}                                  & \multicolumn{1}{l|}{0.500}           & \multicolumn{1}{l|}{0.829}          & \multicolumn{1}{l|}{0.906}          \\ \hline

\multicolumn{1}{|l|}{/}                                                                                  & \multicolumn{1}{l|}{ViT}                       & \multicolumn{1}{l|}{/}                       & \multicolumn{1}{l|}{I}                                  & \multicolumn{1}{l|}{0.394}          & \multicolumn{1}{l|}{0.729}          & \multicolumn{1}{l|}{0.847}          \\ \hline
\multicolumn{1}{|l|}{/}                                                                                  & \multicolumn{1}{l|}{RestNet50}                 & \multicolumn{1}{l|}{/}                       & \multicolumn{1}{l|}{I}                                  & \multicolumn{1}{l|}{0.388}          & \multicolumn{1}{l|}{0.688}          & \multicolumn{1}{l|}{0.806}          \\ \hline
\multicolumn{1}{|l|}{/}                                                                                  & \multicolumn{1}{l|}{CLIP}                      & \multicolumn{1}{l|}{/}                       & \multicolumn{1}{l|}{I}                                  & \multicolumn{1}{l|}{0.482}          & \multicolumn{1}{l|}{0.806}          & \multicolumn{1}{l|}{0.894}          \\ \hline
\end{tabular}
\caption{Top-1, Top-3, and Top-5 accuracy of the model according to fusion methods, encoders, and modalities of the dataset used.}
\label{tab:result}
\end{table}

\section{Conclusion} \label{sec:conclusion}
In this work, we focused on enhancing HS code prediction by leveraging multimodal auxiliary information. We conducted several experiments by varying the fusion methods and the image encoders to find the optimal combination. Experiments demonstrated that Resnet50 yields better results with a top-3 and top-5 accuracy of 93.5\% and 98.2\% respectively. In addition, we proposed our MultConcat fusion method that performed better than simple concatenation and LMF \cite{liu2018efficient} methods in all trials. The results underscore the effectiveness of using multimodal data for HS code predictions, as it outperforms unimodal solutions by 8.2\% in top-1 accuracy. A possible extension to this work could involve quantifying modality contributions using explainability techniques, as well as developing a fusion method capable of handling missing modalities.


\begin{footnotesize}
\bibliographystyle{unsrt}
\bibliography{06_bibliography}
\end{footnotesize}


\end{document}